\documentclass[rmp,aps,twocolumn,floatfix,groupedaddress,superscriptaddress,amsmath,a4paper,twoside,showkeys]{revtex4}
\usepackage{graphicx}
\usepackage{chemarr}
\usepackage{times}
\usepackage{hyperref}
\usepackage{version}
\usepackage{color}
\graphicspath{{figures/},{figures/pdf/},{figures/eps/}}

\bibpunct{[}{]}{,}{n}{}{,}

\newcommand{\fig}{Fig.}
\newcommand{\eq}{Eq.}

\newcommand{\eqs}{Eqs.}

\newcommand{\cF}{{\mathcal{F}}}
\newcommand{\cO}{{\mathcal{O}}}

\providecommand{\avg}[1]{\left \langle #1 \right \rangle}
\providecommand{\pnt}[1]{\left ( #1 \right)}

\providecommand{\addhyphen}[1]{#1.---}

\makeatletter
\renewcommand \paragraph{%
  \@startsection
    {paragraph}%
    {4}%
    {\parindent}%
    {\z@}%
    {-.1em}%
    {\normalfont\normalsize\itshape\addhyphen}%
}%
\makeatother

\begin{document}

\title{
Switching and growth for microbial populations in catastrophic
  responsive environments}

\author{Paolo Visco}

\affiliation{SUPA, School of Physics and Astronomy, The University of
  Edinburgh, The King's Buildings, Mayfield Road, Edinburgh EH9 3JZ,
  UK}

\author{Rosalind J. Allen}

\affiliation{SUPA, School of Physics and Astronomy, The University of
  Edinburgh, The King's Buildings, Mayfield Road, Edinburgh EH9 3JZ,
  UK}

\author{Satya N. Majumdar}

\affiliation{Laboratoire de Physique Th\'eorique et Mod\`eles
  Statistiques (UMR 8626 du CNRS), Universit\'e Paris-Sud, B\^atiment
  100 91405 Orsay Cedex, France}

\author{Martin R. Evans}

\affiliation{SUPA, School of Physics and Astronomy, The University of
  Edinburgh, The King's Buildings, Mayfield Road, Edinburgh EH9 3JZ,
  UK}

\date{\today}

\begin{abstract}
Phase variation, or stochastic switching between alternative states of
gene expression, is common among microbes, and may be important in
coping with changing environments. We use a theoretical model to
assess whether such switching is a good strategy for growth in
environments with occasional catastrophic events. We find that
switching can be advantageous, but only when the environment is
responsive to the microbial population. In our model, microbes switch
randomly between two phenotypic states, with different growth
rates. The environment undergoes sudden ``catastrophes'', the
probability of which depends on the composition of the population. We
derive a simple analytical result for the population growth rate. For
a responsive environment, two alternative strategies emerge. In the
``no switching'' strategy, the population maximises its instantaneous
growth rate, regardless of catastrophes. In the ``switching''
strategy, the microbial switching rate is tuned to minimise the
environmental response. Which of these strategies is most favourable
depends on the parameters of the model.  Previous studies have shown
that microbial switching can be favourable when the environment
changes in an unresponsive fashion between several states. Here, we
demonstrate an alternative role for phase variation in allowing
microbes to maximise their growth in catastrophic responsive
environments.
\end{abstract}


\maketitle
\section{Introduction}

Microbial cells often exhibit reversible stochastic switching between
alternative phenotypic states, resulting in a heterogeneous
population. This is known as phase variation
\cite{woude2004,woude2006,henderson1999}.  A variety of molecular
mechanisms can lead to phase variation, including DNA inversion, DNA
methylation and slipped strand mispairing
\cite{woude2004,woude2006}. These are generally two-state systems
without any underlying multistability \cite{VAE,VAE2}; however,
bistable genetic regulatory networks can also lead to stochastic
phenotypic switching \cite{ptashne,novick,carrier,warren}. The
biological function of phase variation remains unclear, but it has
been suggested that it can allow microbes to evade host immune
responses, or access a wider range of host cell receptors
\cite{henderson1999,hallet2001}. Theoretical work has focused on phase
variation as a mechanism for coping with environmental changes.
According to this hypothesis, a fraction of the population is
maintained in a state which is currently less favourable, but which
acts as an ``insurance policy'' against future environmental changes
\cite{seger}.

In this paper, we present a theoretical model for switching cells
growing in an environment which occasionally makes sudden attacks on
the microbial population.  Viewing the situation from the perspective
of the microbes, we term these {\em{catastrophes}}. These catastrophes
affect only one phenotypic state. Importantly, the environment is
{\em{responsive}}: the catastrophe rate depends on the microbial
population. By solving the model analytically, we find that there are
two favoured tactics for microbial populations in environments with a
given ``feedback function'': keep all the population in the fast
growing state, regardless of the environmental response, or
alternatively, use switching to maintain a population balance that
reduces the likelihood of an environmental response.  Which of these
strategies is optimal depends on the parameters of the model. In the
absence of any feedback between the population and environment, phase
variation is always unfavourable. However, as the environment becomes
more responsive  switching can be advantageous.

Previous theoretical studies have considered models in which the
environment flips randomly or periodically between several different
states, each favouring a particular phenotype.  The case of two
environmental states and two phenotypes has been well-studied
\cite{lachmann,ishii,thattai,gander,ribeiro,wolf1}.  This work has
shown that the total growth rate of the population can be enhanced by
phenotypic switching (compared to no switching), for some parameter
regimes, and that the optimum switching rate is tuned to the
environmental flipping rate. Several studies have also compared random
switching to a strategy where cells detect and respond to
environmental changes. Wolf {\em{et al.}} \cite{wolf1} used
simulations to show that in this case the advantage of random
switching depends on the accuracy of environmental sensing, while in a
theoretical study Kussell and Leibler~\cite{kussell} showed that the
advantages of random switching depend on the cost of environmental
sensing for a model with $n$ phenotypic states and $n$ different
environments. The predictions of the ``two environment, two phenotypic
state'' model have recently been verified experimentally with a
tunable genetic switch in the yeast {\em{Saccharomyces cerevisiae}}
\cite{acar2}.

Here, we consider a different scenario to the above-mentioned body of
work. Rather than considering multiple environmental states, our model
has a single environment, which undergoes occasional, sudden and
instantaneous catastrophes. We assume that the more slowly-growing
microbial phenotypic state is resistant to these
catastrophes. Catastrophic events are likely to be a common feature of
microbial population dynamics in nature. For example, microbes
infecting an animal host may be subject to sudden ``flushing'' due to
diarrhoea or urination, to which they may be resistant if they are
able to attach to the wall of the host's intestinal or urinary tract.
Another example of a catastrophe might be sudden exposure of a
population to antibiotics: here, cells that are in the non-growing
``persister'' state survive, while others are killed
\cite{balaban,kussell2}.  We do not, however, aim to model a specific
biological case, but rather to construct a generic model leading to
general conclusions.

Importantly, and in contrast to previous models, we include in our
model {\em{feedback}} between the microbial population and the
environment: the probability of a catastrophe depends on the state of
the population. Although our model is very general, many examples
exist in nature in which environmental responses are triggered by
characteristics of a growing microbial population, the most obvious
perhaps being a host immune response \cite{mulvey2002}.  Our work
leads us to propose an alternative possible role for phase variation,
to our knowledge not considered in previous theoretical work: we find
that in responsive catastrophic environments, switching can allow the
population to maximise its growth rate while minimising the
environmental response.

The paper is organised as follows. In section \ref{sec:model}, we
present our model. We derive an analytical result for the steady-state
statistics of the model in Section \ref{sec:ss}, and we use this to
predict the optimal strategies for microbial growth, as a function of
the model parameters, in Section \ref{sec:opt}. Finally, we present
our conclusions in Section \ref{sec:conc}.

\section{Model}\label{sec:model}
We consider two microbial sub-populations $A$ and $B$, representing
two different phenotypic states. Between catastrophes, microbes in
these sub-populations grow exponentially at rates $\gamma_A$ and
$\gamma_B$, and switch between states with rates $k_A$ and $k_B$ ($A$
to $B$ and $B$ to $A$ respectively). However, this growing regime can
be ended suddenly by a catastrophe, which consists of a sharp decrease
in the size of the $A$ sub-population. After the catastrophe, the
population dynamics restarts.
 
Between catastrophes the dynamics of the numbers of microbes $n_A$ and
$n_B$ in the two sub-populations are defined by the following system
of differential equations
\begin{subequations}
\label{eq:evol_n}
\begin{equation}
\frac{d n_A}{d t}= \gamma_A n_A + k_B n_B - k_A n_A
\,\,,
\end{equation}
\begin{equation}
\frac{d n_B}{d t}= \gamma_B n_B+ k_A n_A - k_B n_B \,\,.
\end{equation}
\end{subequations}
This description assumes that the population sizes $n_A$ and $n_B$ are
large enough to be considered as continuous variables.  We assume that
$\gamma_A > \gamma_B$, which means that the $A$ sub-population
proliferates faster than the $B$ sub-population.

Whenever a catastrophe takes place, the population size $n_A$ drops
instantaneously to some new value $n_A'<n_A$, with a probability
$\nu(n_A'|n_A)$. The rate at which catastrophes happen depends on the
population size through an {\em environmental response function}
$\beta(n_A, n_B)$. This function characterises the rate at which the
environment responds to the growing population.  The two functions
$\beta$ and $\nu$ are discussed in detail in Section
\ref{catastrophes}.  A typical trajectory for the sizes of the $A$ and
$B$ sub-populations, for a particular choice of $\beta$ and $\nu$, is
shown in the top panel of \fig \ref{fig:sample}.

\subsection{Fitness}

As shown in \cite{thattai}, the two-variable system defined by
Eqs. (\ref{eq:evol_n}) can be replaced by a non-linear dynamical
equation for a single variable.  This variable, $f$, is the fraction
of the total population in the $A$ state:
\begin{equation}
f(t)=\frac{n_A}{n_A+n_B}\,\,.
\end{equation}

If we consider the dynamics of the 
total population $n(t)=n_A(t)+ n_B(t)$, then, from (\ref{eq:evol_n})
it follows that~\cite{thattai}:
\begin{equation}\label{eq:lab11}
\frac{d n(t)}{d t} = \gamma_A n_A + \gamma_B n_B=(\gamma_B + \Delta
\gamma f) n(t) \,\,,
\end{equation}
where $\Delta \gamma=\gamma_A-\gamma_B >0$. The above equation shows
that $f$ is linearly related to the instantaneous growth rate of the
population (which is given by $\gamma_B + \Delta \gamma f$). For this
reason, and following \cite{thattai}, we refer to $f$ as the {\em
  population fitness}.

\begin{figure}
\includegraphics[width=0.95 \columnwidth]{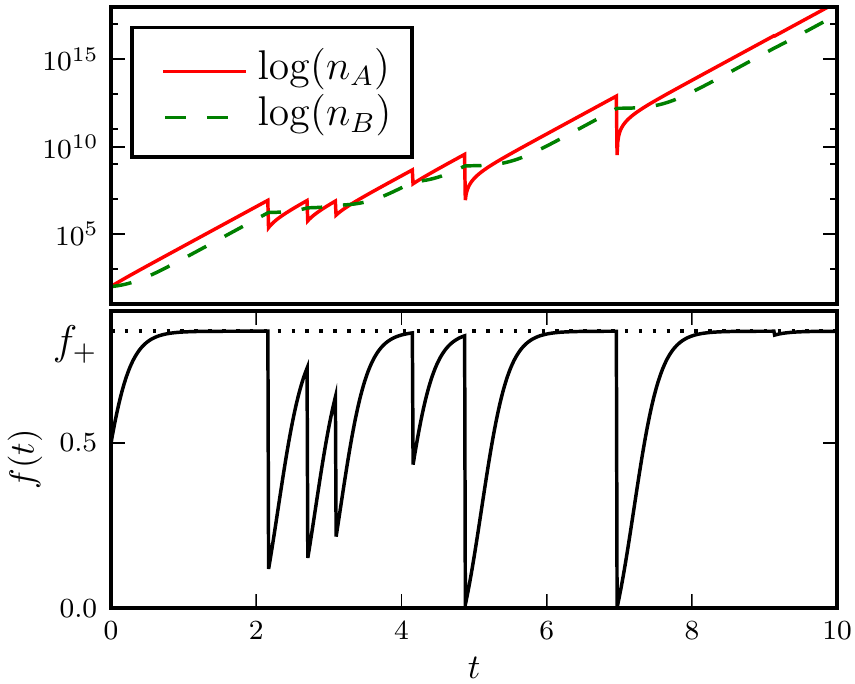}
\caption{\label{fig:sample} Typical trajectory of the system. Top:
time evolution of the populations $n_A$ and $n_B$ (semi-log
scale). Bottom: Corresponding time evolution of the fitness $f(t)$
(the fraction of cells in the $A$ state). Parameters are:
$(\gamma_A,\gamma_B,k_A,k_B,f^*,\beta_0,\alpha)=(6,1,1,1,0.25,1,-0.75)$.}
\end{figure}

The dynamical equation for the population fitness can be determined
from \eqs (\ref{eq:evol_n}) and corresponds to:
\begin{equation}
\label{eq:fitness}
\frac{d f}{d t} = v(f) = - \Delta \gamma (f-f_+)
(f-f_-)\,\,,
\end{equation}
where we define $v(f)$ as the time evolution function for the fitness,
and $f_\pm$ are the two roots of the quadratic equation:
\begin{equation}\label{eq:roots}
f^2 -\left(1- \frac{k_A + k_B}{\Delta \gamma} \right) f -
\frac{k_B}{\Delta \gamma} =0 \,\,.
\end{equation}
One can check that the smaller root takes values $f_-<0$, while the
larger root takes values $0<f_+\le 1$. Hence, the population fitness
increases towards a plateau value $f_+$, until a catastrophe happens,
upon which it is reset to a lower value. A typical time trajectory for
the population fitness is plotted in the bottom panel of \fig
\ref{fig:sample}. The time evolution of $f$ is deterministic except at
some specific time points (catastrophes) where it undergoes random
jumps. This model can therefore be considered to be a \emph{Piecewise
Deterministic Markov Process} \cite{davis,otto}.

\subsection{Catastrophes}
\label{catastrophes}

The catastrophes in our model have two characteristics: the rate at
which they happen and their strength ({\it i.e.} how many microbes are
killed).  The rate at which catastrophes occur, or their probability
per unit time, is defined by a feedback function $\beta(f)$, which we
take to depend only on the fitness of the population and not on the
absolute population size (we shall return to this assumption in the
Discussion).  The function $\beta(f)$ characterises the response of
the environment to the growth of the population. When $\beta=0$, there
are no catastrophes and the fitness will reach the plateau value $f_+$
and stay there for ever. Nonzero constant values of $\beta$ correspond
to a non-responsive environment in which the catastrophes follow
Poisson statistics.  We shall consider the case of a responsive
environment characterised by a response function $\beta(f)$ which
depends on the population fitness. In particular, we consider a
nonlinear response function that has a sigmoid shape. Thus, the
probability per unit time of a catastrophe is very low when the
population fitness is low, but increases significantly if the fitness
exceeds some threshold value. This scenario might correspond to a
detection threshold in the environment's sensitivity to population
growth.

The precise environmental response function that we consider is the
following:
\begin{equation}
\beta_\lambda(f)=\frac{\xi}{2} \pnt{1 +
\frac{f-f^*}{\sqrt{\lambda^2 + (f-f^*)^2}}}\,\,.
\label{lambda}
\end{equation}
\begin{figure}
\includegraphics[width=0.95 \columnwidth]{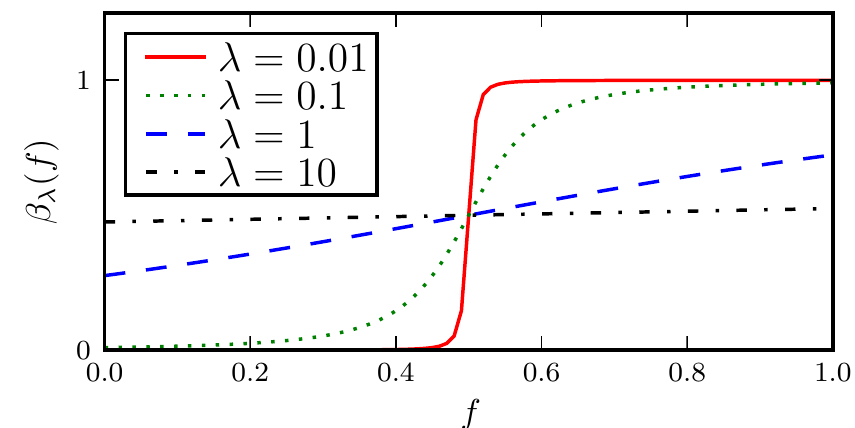}
\caption{\label{fig:beta} The response function $\beta_\lambda(f)$,
plotted for different values of $\lambda$, with $\xi=1$ and
$f^*=1/2$. }
\end{figure}
Although this function is defined for all $-\infty < f <
\infty$, the relevant interval for the fitness is $0 < f < 1$.
Typical shapes for this function are shown in \fig \ref{fig:beta}.
The parameter $\xi$ is the asymptotic value of $\beta_\lambda$ when $f$
is large, and we refer to $\xi$ as the saturated catastrophe rate.
As the population fitness $f$ increases, $\beta_\lambda$ increases
from 0 to $\xi$ around the threshold value $f^*$ at which
$\beta_\lambda = \xi/2$.  Finally, the parameter $\lambda$ determines
the sharpness of the threshold. For small values of $\lambda$ the
function $\beta_\lambda(f)$ approaches a step function
\begin{equation}
\beta_0(f)= \xi \Theta(f-f^*)\,\,.
\end{equation}
As $\lambda$ increases, the function  broadens and becomes linear 
over a range of $f$ near $f^*$:
\begin{equation}
\beta_\lambda \simeq \frac{\xi}{2} \pnt{1 + \frac{(f-f^*)}{\lambda} +
\cO\pnt{\frac{1}{\lambda^2}}}\,\,.
\end{equation}
while when the parameter $\lambda$ becomes very large ($\lambda \to
\infty$), $\beta_\lambda(f)$ becomes constant (independent of $f$) so
that the catastrophes become a standard Poisson process with parameter
$\xi/2$:
\begin{equation}
\beta_\infty(f)= \xi/2 \,\,.
\label{llambda}
\end{equation}
We emphasise that we have chosen this particular sigmoid function
(\ref{lambda}) as the $\lambda$ parameter allows a convenient tuning
of its shape and thus the degree of environmental responsiveness.  
However, our conclusions are not affected by the
particular choice of sigmoid function.

We now turn to the function describing the catastrophe strength,
$\nu(n_A'|n_A)$. This is the probability that, given that $n_A$ cells
of type $A$ are present before the catastrophe, $n_A'$ will remain
after the catastrophe. In order to retain our description of the model
in terms of the population fitness, we shall consider that $\nu$ only
depends on $n_A'$ only through the ratio $n_A'/n_A$.  Then the
normalisation of $\nu$ implies that
\begin{equation}
\nu(n_A'|n_A)=\frac{1}{n_A} \cF(n_A'/n_A)
\label{nuF}
\end{equation}
where $\int_0^1 dx \cF(x) =1$.  When a catastrophe happens, the
population size is reduced by a random factor sampled from the
distribution $\cF$ [{\it i.e.} the new size $n_A'=n_A \times u$ where
$n_A$ is the size before the catastrophe, and $u$ is a random number
($0 \le u <1$) sampled from the distribution $\cF$]. This allows us to
associate to each jump $n_A \to n_A'$ a fitness jump $f \to f'$, where
$f'=n_A'/(n_A'+n_B)$. The size of these jumps will be distributed
according to:
\begin{equation}\label{eq:lab22}
\mu(f'|f)= \Theta\pnt{f -f'} \cF\pnt{\frac{f' (1-f)}{f (1-f')}}
\frac{1-f}{(1-f')^2 f }\,\,.
\end{equation}
Eq.(\ref{eq:lab22}) can be obtained by rewriting expression
(\ref{nuF}) for $\nu(n_A'|n_A)$ as a function of $f$ and $f'$ and
including the Jacobian of the transformation.

\begin{figure}
\includegraphics[width=0.95 \columnwidth]{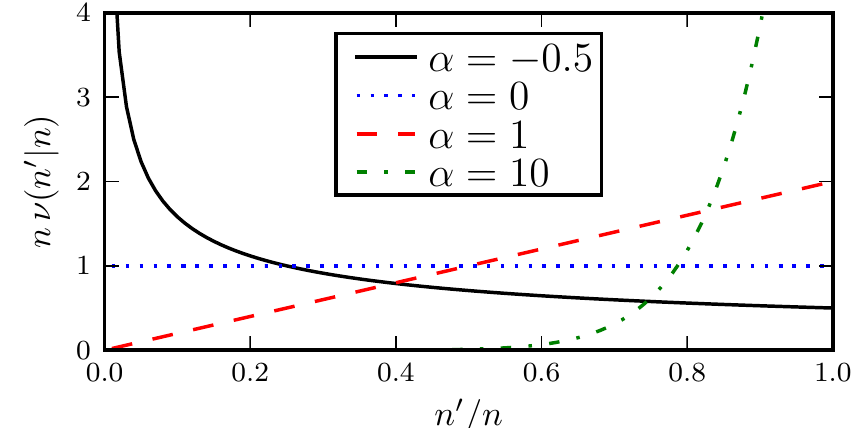}
\caption{\label{fig:nu}The jump distribution $\nu(n_A'|n_A)$, plotted
  for various values of $\alpha$.}
\end{figure}

In this paper, we shall consider the simple case where $\cF(x)=
(\alpha+1) x^\alpha$, with $\alpha>-1$. 
The explicit expression for $\nu(n'|n)$ thus reads:
\begin{equation}
\nu(n'|n)= \frac{(\alpha +1)}{n} \pnt{\frac{n'}{n}}^\alpha \,\,, \qquad \qquad
\alpha>-1\,\,.
\label{nua}
\end{equation}
This choice is made primarily in order to allow us to solve the model
analytically: the choice implies that $\mu(f'|f)$ factorises (see
(\ref{eq:form})) which then allows the integral equation for the
probability flux balance (\ref{eq:prflux}) to be solved.  Moreover,
the choice of a power law distribution for $\nu(n_A'| n_A)$ is general
in that allows for increasing, decreasing or flat functional forms.
The function $\nu(n_A'|n_A)$ is plotted in \fig \ref{fig:nu} for
various values of $\alpha$. For negative $\alpha$ values, the
distribution is biased towards far-reaching catastrophes that reduce
fitness significantly. The case $\alpha=0$ corresponds to jumps
sampled from a uniform distribution, whereas positive values give a
distribution biased towards weaker catastrophes.  The parameter
$\alpha$ can therefore be used to tune the strength of the
catastrophes (although in this work we shall always consider negative
$\alpha$ values, corresponding to strong catastrophes).
We note here that with our choice for $\cF(x)$ the jump distribution
can be expressed as
\begin{equation}
\label{eq:form}
\mu(f'|f)=\Theta(f-f') \frac{d}{df'} \frac{m(f')}{m(f)}\,\,.
\end{equation}
where $m(0) =0$, $\int df' \mu(f'|f)=1$, and with,
\begin{equation}
\label{eq:m}
m(f)= \pnt{\frac{f}{1-f}}^{1+\alpha} \,\,.
\end{equation} 
%However, we will see below that analytical solution can be obtained
%for any jump distribution of the form (\ref{eq:form}) with an
%arbitrary function $m(f)$ which is not necessarily of the form
%(\ref{eq:m}).

\section{Steady--state statistics}\label{sec:ss}

\begin{figure}
\includegraphics{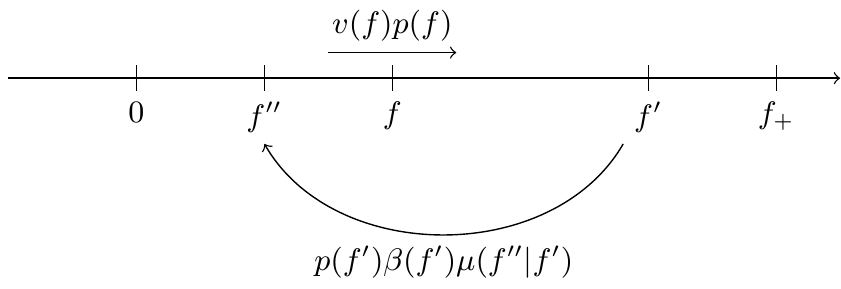}
\caption{\label{fig:balance} Illustration of the flux balance
  condition. The positive probability flux due to population growth
  must be balanced by the negative flux due to catastrophes.}
\end{figure}

We now derive the steady-state probability distribution for the
population fitness, $p(f)$.  The distribution $p(f)$ must satisfy a
condition of balance for the probability flux. This condition reads:
\begin{equation}
\label{eq:prflux}
v(f)p(f) = \int_f^{f_+} df' \int_0^f  df'' \beta(f') p(f') \mu(f''|f')
 \,\,.
\end{equation}
The left-hand side of the above equation corresponds to the
deterministic probability flux due to population growth as defined in
Eq.(\ref{eq:fitness}) [$f(t)$ increases in time as the population
grows, as shown in Figure \ref{fig:sample}]. The right-hand side
describes the probability flux arising from catastrophes. In this
model, catastrophes always reduce the population
fitness. The probability flux due to catastrophes therefore contains
contributions from all possible jumps that start at some $f'>f$ and
end at some $f''<f$. These contributions must be weighted by
$\beta(f')p(f')$: the probability of having fitness $f'$ {\em and}
undergoing a catastrophe. This balance between the fluxes due to
growth and catastrophes is illustrated schematically in Figure
\ref{fig:balance}.

Inserting the form (\ref{eq:form}) of $\mu(f'|f)$ in
Eq.(\ref{eq:prflux}), the zero flux condition becomes:
\begin{equation}
v(f) p(f)= \int_f^{f_+} df' \beta(f') p(f') \frac{m(f)}{m(f')}\,\,.
\end{equation}
We now divide the above equation by $m(f)$ and take the first
derivative with respect to $f$. This yields, in terms of the function
$G=vp/m$:
\begin{equation}
\frac{d G}{df}= - \frac{\beta G}{v}\,\,.
\end{equation}
The above differential equation is then easily solved for $G$. The
result for $p(f)$, using the form \eq (\ref{eq:m}) of $m(f)$ , is
finally
\begin{equation}
\label{eq:sol} p(f)=\frac{C}{v(f)} \pnt{\frac{f}{1-f}}^{1+\alpha} \exp
\pnt{-\int \,df\, \frac{\beta(f)}{v(f)}}\,\,,
\end{equation}
where $C$ is a normalisation constant.  The integral in
Eq.(\ref{eq:sol}) can be performed analytically for the model defined
in the previous section. The result, which is rather cumbersome, is
given in the Appendix. 

We present in \fig\ref{fig:samplepf} (top panel) 
some resulting shapes for the probability distribution $p(f)$
in the case $\lambda = 0$, corresponding to a step function for the environmental response.
We consider two different values of $\Delta \gamma$,
in each case for  $k_A =0$ (no switching) and  a  non-zero switching
rate $k_A= k_A*$  defined such that $f_+ = f^*$ (see section 
\ref{sec:opt}).
In these plots we see that singularities in $p(f)$ can arise
at $f= 0,f^*$ or $f_+$ in different cases.
In particular, the $k_A=0$ case produces a cusp at $f=f^*$
(due to the singular nature of the
step function $\beta(f)$ at $f^*$)
for small $\Delta \gamma$,  and/or a divergence at 
$f=f_+$ for  large $\Delta \gamma$. On the other hand
$k_A= k_A*$ produces a divergence at $f_+ = f^*$ in both cases.

\fig\ref{fig:samplepf} (bottom panel) plots trajectories of the fitness
corresponding to the parameter values of \fig\ref{fig:samplepf}.
These trajectories reveal the interplay between two timescales:
the time to relax to the plateau value $f^*$ in the absence of
catastophes and the typical time between catastrophes. The former
decreases with $\Delta \gamma$ and the latter is given by $1/\xi$
where $\xi$ is the plateau value of the response function $\beta$. A
divergence of $p(f)$ at $f= f_+$ arises when the plateau value is
typically reached before a catastrophe occurs.

%Figures \ref{fig:samplepf}(a)
%and (b) show smooth peaks at fitness values intermediate between $f=0$
%and $f_+$, while \fig \ref{fig:samplepf} (c) illustrates a cusp which
%appears at the threshold value $f=f^*$ in the limit $\lambda \to 0$
%when $\beta(f)$ tends to a step function. Finally, \fig
%\ref{fig:samplepf} (d) shows a bimodal-like distribution that can be
%obtained in a regime where $k_B$ is small while $\Delta \gamma$ is
%large.
\begin{figure}
\includegraphics[width=0.95 \columnwidth]{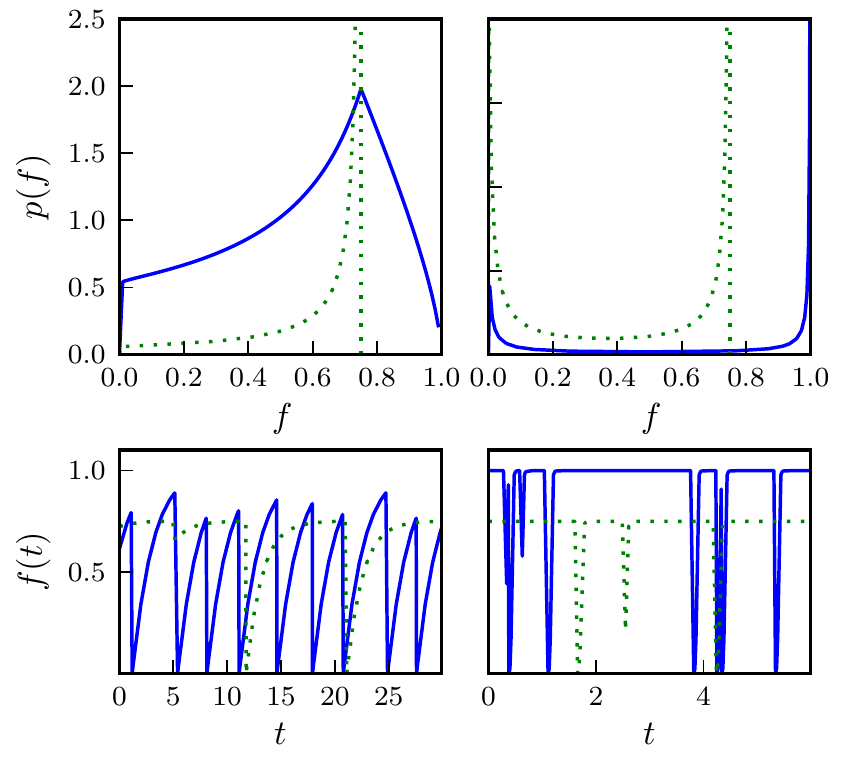}
\caption{\label{fig:samplepf} Top panel: Examples of steady-state fitness
    distribution $p(f)$
    for parameter values  $\Delta \gamma$= 0.1 (left) and 100 (right); other
    parameters are $\lambda=0$, $k_B=0.5$, $\xi=1$, $\alpha=-0.99$ and
    $f^*=0.75$. In each plot, the solid line corresponds to a
    value of $k_A=0$ (no switching), while the dashed plot
    is for $k_A=k_A^*$ (switching rate given by Equation \ref{ka*}). 
    Bottom panel: Examples of fitness trajectories corresponding to the parameter
values of the top panel.
}
\end{figure}

\section{Optimal strategies: to switch or not to switch?}\label{sec:opt}
The key question to be addressed in this work is whether random
switching is advantageous to the microbial population in our model. To
answer this question, we take advantage of the analytical solution
(\ref{eq:sol}) to investigate how the time-averaged population fitness
depends on the rate $k_A$ of switching from the fast-growing state $A$
to the slow-growing state $B$. We are particularly interested in the
effect of the parameter $\lambda$, which controls the sharpness of the
environment's response to the population. 

\begin{figure}
\includegraphics[width= 0.95 \columnwidth]{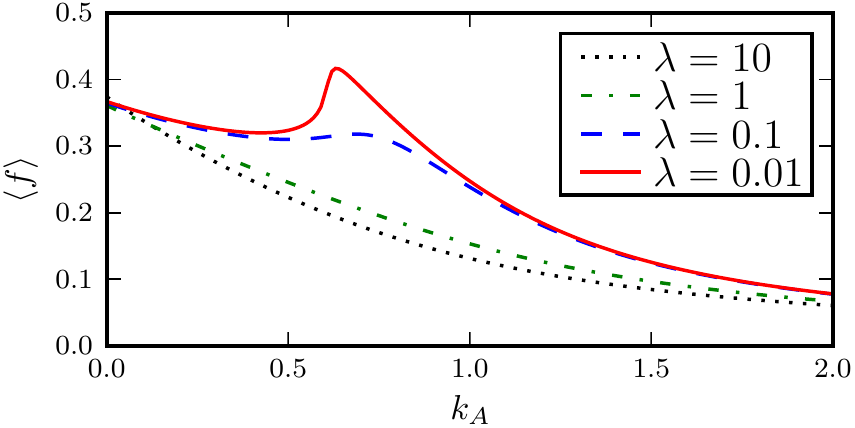}
\caption{\label{fig:ave_lambda} Average population fitness $\avg{f}$
  as a function of the $A \to B$ switching rate $k_A$, for several
  values of parameter $\lambda$. For large values of $\lambda$ the
  average fitness peaks for zero switching, $k_A=0$. However, as
  $\lambda$ decreases, a second peak arise at non zero $k_A$,
  indicating that random switching can be a favourable strategy when
  the environment is responsive. Parameters values are: $k_B=0.1$,
  $\Delta \gamma=1$, $\alpha=-0.99$, $\xi=1$ and $f^*=0.5$.}
\end{figure}

In \fig \ref{fig:ave_lambda} we plot the average population fitness
against $k_A$ for several values of $\lambda$. For a non responsive
environment [i.e. in the limit of large $\lambda$, where the
catastrophe rate takes the constant non-zero value $\xi/2$
(\ref{llambda})], the population fitness has only one (boundary)
maximum for switching rate $k_A \to 0$. This means that the optimal
rate of population growth is achieved when the bacteria do not switch
away from the fittest state $A$.  It should be noted that we consider
the limit $k_A \to 0$, so that the population always contains some
small residual fraction in the unfit $B$ state, which becomes a finite
fraction of the population after a catastrophe.  Subsequently, in
between the catastrophes, the $A$ subpopulation grows quickly to
dominate the population and the fitness evolves towards the value
$f_+=1$, which follows from \eq (\ref{eq:roots}) when $k_A=0$.

 In contrast, as the environment is made responsive by decreasing the
parameter $\lambda$, a local maximum appears in the population
fitness, for nonzero switching rate $k_A$. This implies that for
responsive environments, switching into the slow-growing state
represents an optimal strategy for the microbes. The height of the
peak at $k_A \ne 0$ can surpass that of the peak at $k_A=0$, showing
that random switching can be advantageous compared to keeping the
whole population in the fast-growing state, if the environment is
responsive. Thus the two maxima correspond to two alternative
strategies which we term {\em switching} for the peak at $k_A=k_A^*$
and {\em non-switching} for the peak at $k_A=0$.

To gain further insight into the meaning of these two strategies, and
to determine which circumstances favour one strategy over the other,
we  focus on the limiting case $\lambda=0$, where the response
function is a step function with its threshold at $f=f^*$. We 
assume that the environmental threshold $f^*$ is less than the maximum
population fitness $f_+$. (If this is not the case, the unrealistic
situation arises where the population never has a high enough fraction
of $A$ cells to trigger any catastrophes.) Since $f_+$ depends on the
switching rate $k_A$ via Eq.(\ref{eq:roots}), this condition $f^*<f_+$
implies a maximum value $k_A^*$ for $k_A$:
\begin{equation}
k_A^*=\frac{(1- f^*) (\Delta \gamma \, f^* + k_B)}{f^*}\,\,.
\label{ka*}
\end{equation}
\begin{figure}
\includegraphics[width= 0.95 \columnwidth]{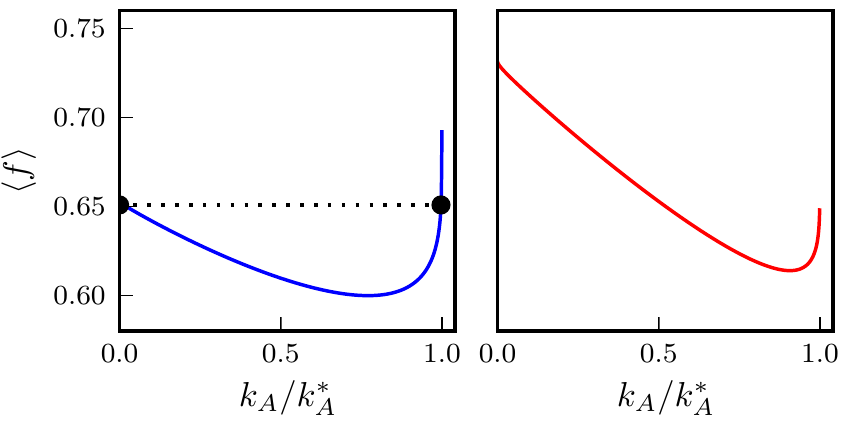}
\caption{\label{fig:kaopt}Average population fitness $\avg{f}$ as a
  function of $k_A$ for a threshold response function. The left plot
  is an instance where the maximal fitness occurs at non-zero $k_A$,
  while the right plot shows an example where the maximum fitness
  occurs at $k_A=0$. In both plots parameters are $\alpha=-0.99$,
  $f^*=0.75$, $\beta_0=1$ and $k_B=0.8$; $\Delta \gamma=1$ (left) and
  $\Delta \gamma=4$ (right).}
\end{figure}
\fig \ref{fig:kaopt} shows two examples of how the average fitness
$\avg{f}$ depends on $k_A$ in the range 0 to $k_A^*$. One can see that
there are always two boundary maxima located at $k_A=0$ and at
$k_A=k_A^*$; these correspond to the non-switching and switching
strategies.

\begin{figure}
\includegraphics[width=0.95 \columnwidth]{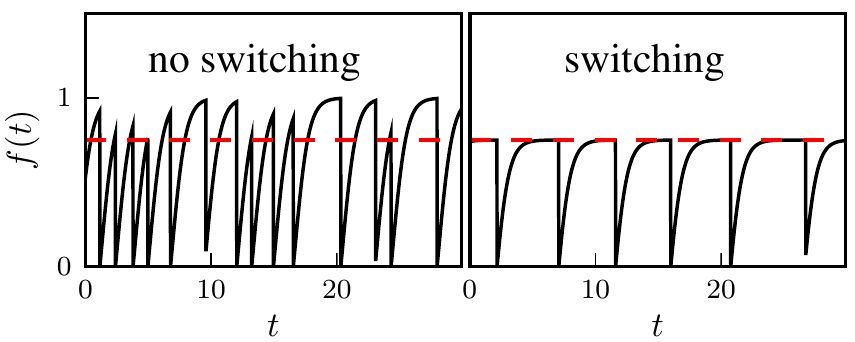}
\caption{\label{fig:two_strategies} Two typical trajectories
  illustrating the two possible strategies: in the left panel, the
  switching rate $k_A=0$, the fitness evolves towards its maximal
  value $f=1$ but many catastrophes are induced; in the right panel
  $k_A$ is set to a value smaller but very close to $k_A^*$
  ($k_A=0.6155=0.997 k_A^*$), the fitness evolves to just above the
  threshold value with relatively few catastrophes induced. These
  strategies correspond to the two black points in the top left plot
  of \fig \ref{fig:kaopt}. The value of $f^*$ is 0.75, as indicated by
  the red dashed line, and the average fitness is the same in both
  plots ($\avg{f}=0.652585$).}
\end{figure}

We plot in \fig \ref{fig:two_strategies} typical trajectories of the
population fitness for the two cases corresponding to the black
circles in \fig \ref{fig:kaopt} (left panel). These trajectories have
the same time-averaged population fitness, but they show very
different dynamical behaviour. The non-switching strategy ($k_A=0$) is
characterised by a fast evolution of the fitness towards its maximum
$f_+=1$. However this triggers frequent catastrophes which cause
sudden decreases in fitness. In contrast, for the switching strategy
($k_A \to k_A^*$), the fitness has a slower growth towards a plateau
value at the detection threshold $f^*$. In this way, the population
reduces the frequency of catastrophes by maintaining itself in a
heterogeneous state with a non-zero fraction of slower-growing cells
which do not trigger catastrophes.

\begin{figure}
\includegraphics[width=0.95 \columnwidth]{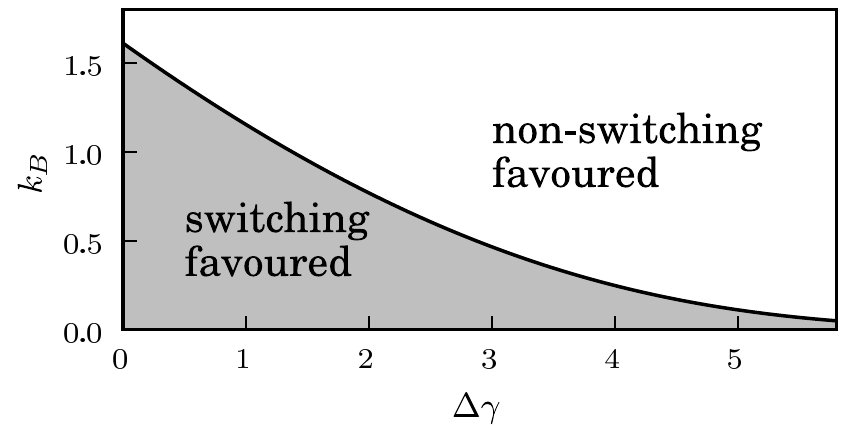}
\caption{\label{fig:phasediagram1}``Phase diagram'' showing the
  optimal strategy as a function of the parameters $k_B$ and $\Delta
  \gamma$ which describe the microbial population. In the shaded area
  switching is the optimal strategy, while in the unshaded area, the
  non-switching strategy leads to a larger average population fitness.
  The other parameters are: $\alpha=-0.99$, $f^*=0.75$, $\xi=1$.}
\end{figure}

\begin{figure}
\includegraphics[width=0.95 \columnwidth]{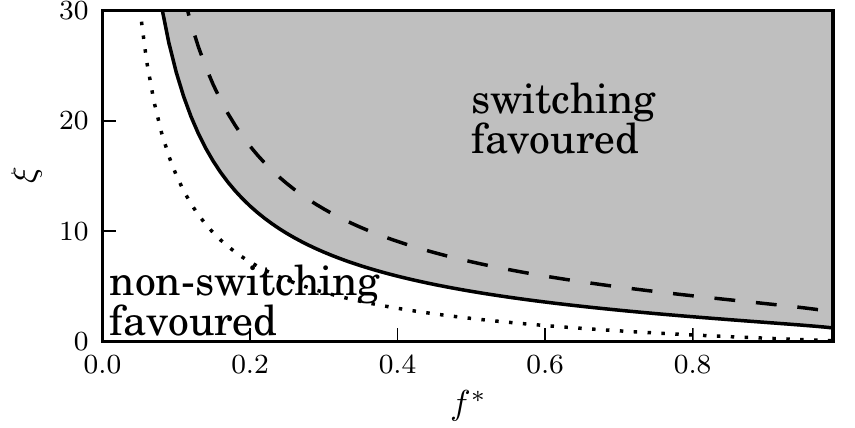}
\caption{\label{fig:phasediagram2} ``Phase diagram'' showing the
 optimal strategy as a function of the parameters $\xi$ and $f^*$
 which describe the environmental response.  In the shaded area
 switching is the optimal strategy, while in the unshaded area, the
 non-switching strategy leads to a larger average population fitness.
 The other parameters are $\Delta \gamma=1$ and $k_B=1$. The black
 solid line shows the ``phase boundary'' for $\alpha=-0.5$, while the
 dotted and dashed line correspond to $\alpha=-0.99$ and $\alpha=0$.}
\end{figure}

We next consider how the parameters of our model affect the balance
between the switching and non-switching strategies. To this end, we
plot ``phase diagrams'' showing which of these two strategies achieves
a higher population growth rate for a given set of parameters.
Figure~\ref{fig:phasediagram1} considers the parameters describing the
microbial population: the difference $\Delta \gamma$ in growth rate
between the $A$ and $B$ states and the switching rate $k_B$ from the
slow-growing $B$ state to the fast-growing $A$ state. This diagram
shows that the switching strategy is only favourable when the $B$
state does not carry too high a cost in terms of growth rate ($\Delta
\gamma$ not too large) and when switching to the $B$ state is unlikely
to be immediately followed by a reverse switch back into the $A$ state
($k_B$ not too large). In \fig \ref{fig:phasediagram2} we consider
instead the parameters describing the environmental response: the
detection threshold $f^*$ and the saturated catastrophe rate
$\xi$. Here, we see that the switching strategy ({\em{i.e.}}
attempting to avoid catastrophes) is favoured when the saturated
catastrophe rate $\xi$ is high or when the threshold value $f^*$ is
high (since for high thresholds the population does not have to pay a
very high price in terms of $B$ cells to avoid triggering
catastrophes). For very low detection thresholds $f^*$, lower than
typical values of the fitness, the environmental response will almost
always ``detect'' the population, and the environmental behaviour will
thus be similar to the situation of a non--responsive environment,
which corresponds to the limiting case where $f^*=0$. In this case, as
discussed earlier, non-switching is the optimal strategy.
Figure~\ref{fig:phasediagram2} also demonstrates the effect of
changing the catastrophe strength parameter $\alpha$ (dashed and
dotted lines). The switching strategy is favoured by strong
catastrophes (negative $\alpha$), while the non-switching strategy is
more likely to be optimal for weak catastrophes (i.e. larger positive
$\alpha$). All this points to the conclusion that in general,
switching tends to be an advantageous strategy when the 
characteristics of the catastrophic
environment  are particularly adverse (large $\xi$ and negative
$\alpha$) {\em and} when the detection threshold is not too low.

\section{Discussion and further directions}\label{sec:conc}

In this work we have considered the possible advantages of phase
variation (random switching between phenotypic states) for a microbial
population in a catastrophic responsive environment. To this end, we
solved analytically for the steady-state statistics of a model which
includes two microbial sub-populations that  grow and switch, and a
single environment which occasionally mounts catastrophic attacks on
the microbial population. Importantly, the model includes feedback
between the state of the population and the frequency of catastrophic
events via an environment response function which depends on the
population through its fitness i.e. the instantaneous rate of
growth. Our results show that,
when the environment is responsive to
the population,  switching can increase the average fitness (i.e. growth rate)
of the population. A general picture emerges from our work of two competing
strategies for dealing with a catastrophic responsive environment:
not switching and thus
maximising the instantaneous growth rate regardless of catastrophes versus
using switching to tune the population to reduce the likelihood of 
catastrophes.

An important feature of this work is the fact that we are able to
solve the model analytically, leading to an explicit formula for the
population fitness as a function of the model parameters. 
%We recall
%that the population fitness is linearly related to the average
%population growth rate following (\ref{eq:lab11}). 
To achieve this analytical result, we make a number of assumptions,
the most important being that the environmental feedback depends on
the instantaneous growth rate rather than on the population size.
Although this is a rather idealized assumption, microbe-host
interactions are in reality likely to be sensitive to microbial growth
rate \cite{johri}, since several intracellular small molecules and
proteins, including ppGpp, cAMP and H-NS, whose concentrations are
growth-rate dependent \cite{ferenci,schaechter}, have been shown to
regulate microbial virulence factors
\cite{pizarro,pesavento,schroder}.

The main conclusion of our work is that phase variation can provide a
mechanism by which a microbial population can tune its composition so
as to  minimize the likely environmental response, thus increasing its average
growth rate (or average fitness). The model thus provides an
alternative scenario for the role of phase variation to those proposed
in other theoretical studies, which we now take the opportunity to
review briefly.

Various works have considered models in which the environment flips
randomly or periodically between several different states, each
favouring a particular cell phenotype. These models do not include 
feedback between the population and the environmental flipping
rate. For the case of two environmental states and two cellular
phenotypes, Lachmann and Jablonka \cite{lachmann} considered a
discrete time model with a periodic environment, while Ishii {\em{et
al}} \cite{ishii} addressed a similar problem but explicitly looked
for the evolutionary stable state. Thattai and Van
Oudenaarden~\cite{thattai} also considered the two-environment,
two-phenotype case, using a continuous time model with Poissonian
switching of the environment, while a detailed analytical treatment of
this case was presented by Gander {\em{et al.}} \cite{gander} and a
simulation study was carried out by Ribeiro \cite{ribeiro} with a more
detailed model of the phenotypic switching mechanism, while Wolf
{\em{et al.}} \cite{wolf1} simulated a model that also included
environmental sensing. These studies showed that the total growth rate
of the population can be enhanced by phenotypic switching (compared to
no switching), for some parameter regimes, and that the optimum
switching rate is tuned to the environmental flipping rate.  A similar
model, but aimed specifically at the case of the persister phenotype,
in which cells grow very slowly but are resistant to antibiotics
\cite{balaban}, was considered by Kussell {\em {et al.}} for a
periodic environment \cite{kussell2}. In this model, the growth rate
of the non-persister phenotype is negative (signifying population
decrease) in the ``antibiotic'' environment.

Several other studies have considered random switching from a
different context: as a means to avoid the need for sensing and
responding to environmental changes, in the case that environmental
sensing is inaccurate, faulty or expensive. In this context, Kussell
and Leibler \cite{kussell} considered theoretically a model with many
environments and many cellular states, where a cost is attached to
sensing environmental changes, while Wolf {\em{et al.}} \cite{wolf1}
simulated a two state-two environment model where sensing was subject
to a variety of possible defects. Both these studies concluded that
random switching can be a good strategy to overcome disadvantages
associated with environmental sensing.

In a somewhat different approach, Wolf {\em{et al.}} \cite{wolf2} used
simulations to study a two state-two environment model in which the
growth rate of the $A$ and $B$ states is frequency
dependent---{\em{i.e.}} a given microbial sub-population grows faster
when its abundance is low. Such ``frequency-dependent selection'' is
well known to promote population heterogeneity; however, Wolf {\em{et
al.}} did not find any advantages for reversible switching as a means
to generate this heterogeneity as opposed to terminal cellular
differentiation. In a sense, the model presented in this paper also
incorporates frequency dependent selection, since catastrophes are
less likely when the $A$ sub-population is small. However, in contrast
to Wolf {\em{et al.}}, we find that reversible switching does play an
important role. If switching in our model were not reversible, there
would be no way for the fast-growing $A$ subpopulation to regenerate
from the surviving $B$ cells after a catastrophe.

Although the majority of theoretical work in this area, including that
presented in this paper, has focused on the interplay between cellular
switching and environmental changes, this is not the only
perspective from which the role of phase variation can be viewed. For
example, an alternative scenario, which does not require a changing
environment, was recently presented by Ackermann {\em{et al.}}
\cite{ackermann}. These authors showed that random switching into a
``self-sacrificing'' phenotypic state can be evolutionarily favoured
if the individuals in that state have on average greater access to
some beneficial resource. This idea raises a number of interesting
questions which we hope to pursue in future research.

Finally we note that the theoretical framework developed in the
present work, although applied here to the case of detrimental
and instantaneous catastrophes, could also be used to model environmental changes more
generally. For example, in the symmetric two state-two environment
model considered by Thattai and van Oudenaarden \cite{thattai}, the
environment flips randomly between two states and these flips are
accompanied by a change in fitness from $f$ to $1-f$.  This could be
incorporated in our theoretical framework by setting the $\beta(f)$ to
a constant value and the jump distribution $\mu(f|f')$ to:
\begin{equation}
\mu(f'|f)=\delta\pnt{f'-(1-f)}\,\,.
\end{equation}
However, such a choice of $\mu(f|f')$ would result in fundamentally
different conclusions to those of the present study, because the
fitness in the Thattai-van Oudenaarden model (and in other similar
models) is not necessarily decreased when the environment changes.  In
fact, if a large fraction of the cells are in the slow-growing state
before the environment flips so that $f<1/2$, then the environmental
change will actually increase the fitness of the population. In
contrast, in the present work, all catastrophes are detrimental and
the advantage of switching lies in avoiding the triggering of an
environmental response.

The present study suggests a number of avenues for further
work. First, it would be useful to check the robustness of the results
to changes in the choice of catastrophe distributions. Here we have
adopted the power law (\ref{nua}) which allows the exact solution of
the model and generates a broad range of catastrophes sizes. Such a
distribution could be justified in the context of an `antibiotic'
environment, as representing the dose-response variability of
antimicrobes \cite{NAD} and variability in the dosage. One could also
explore other distributions such as exponentially distributed
catastrophes or those centred about some particular catastrophe
fraction $f'= a f$ with $a<1$.  It remains to be determined which
choice is most biologically relevant in different contexts.

Another point which also deserves investigation in future work is the
relation between the choice of switching strategy and the variability
in the population fitness.  For example in \fig \ref{fig:samplepf} one
can see that the different strategies give very different widths for
the fitness distribution $p(f)$.  In this work we defined the optimal
strategy as that which gives the maximal average growth of the
population. However, it might also be relevant to include fitness
fluctuations in the criteria for optimality.

It is also important to consider the case where the environmental
response depends on the absolute size of a particular subpopulation.
Here, we expect that the population size may reach a steady state
governed by the balance between growth and catastrophes.  The total
population size could then be maximised either by maximising the
growth rate, regardless of catastrophes, or by tuning the population
composition to avoid triggering catastrophes. We thus expect that the
two strategies identified in the present work will prove to be
relevant to a variety of models. Moreover, we note that the
distinction between models based on growth rate and those based on
population size may vanish for scenarios with constant population size
such as chemostat cultures \cite{Ingraham}. Equally interesting are
the prospects for including spatial effects, such as adhesion to host
surfaces, or transfer between different environmental compartments, in
the model, and for generalising the model to include many different
microbial states, in which case the same theoretical framework could
perhaps be used to describe genetic evolution of microbial populations
in catastrophic responsive environments.

\acknowledgements The authors are grateful to David Gally and Otto
Pulkkinen for useful discussions. R. J. A. was funded by the Royal
Society. This work was supported by EPSRC under grant EP/E030173.

\appendix

\section{Explicit form of $p(f)$}
Below we give the explicit form for the integral appearing in
(\ref{eq:sol}) when $\beta(f)$ is given by (\ref{lambda}):
\begin{widetext}
\begin{multline}
\int df\, \frac{\beta(f)}{v(f)}=\frac{\xi}{2 \Delta \gamma \Delta f}
\log \left\{ \frac{(f-f_-)}{(f_+-f)} \left[\frac{2 \Delta f
((f^*-f)(f^*-f_-) + \lambda^2 + g(f,f^*) g(f_-,f^*))}{(f-f_-)(f^*-f_-)
g(f_-,f^*)} \right]^{\frac{(f^*-f_-)}{g(f_-,f^*)}} \right. \\ \times
\left. \left[\frac{2 \Delta f ((f-f^*)(f_+-f^*) + \lambda^2 + g(f,f^*)
g(f_+,f^*)}{(f_+-f) (f_+-f^*) g(f_+,f^*)}
\right]^{\frac{(f_+-f^*)}{g(f_+,f^*)}} \right\}\,\,,
\end{multline}
\end{widetext}
where 
\begin{equation}
g(a,b)=\sqrt{(a-b)^2 + \lambda^2}\,\,.
\end{equation}
From this result, the explicit expression for the fitness distribution
function $p(f)$ can be easily derived.

\bibliographystyle{h-physrev}
\bibliography{environments.bib}
\end{document}